\documentclass[journal]{IEEEtran}
\usepackage{amsmath}
\usepackage{lipsum}
\usepackage{nomencl}
\usepackage{amsmath,amssymb,url}
\usepackage{multicol}
\usepackage{lipsum}
\usepackage{graphicx}
\usepackage{comment}
\usepackage{multirow}

\usepackage{booktabs,makecell}
\ifCLASSINFOpdf

\else

\fi

\usepackage{filecontents}
\usepackage[noadjust]{cite}

\begin{document}

\title{Synthesize Phasor Measurement Unit Data Using Large-Scale Electric Network Models}

\author{\IEEEauthorblockN{Ti Xu, Hanyue Li, Adam B. Birchfield, Thomas J. Overbye} \\
\IEEEauthorblockA{\textit{Department of Electrical and Computer Engineering} \\
\textit{Texas A\&M University}\\
College Station, United States}


\thanks{This work was supported by the U.S. Department of Energy Advanced Research Projects Agency-Energy (ARPA-E) under the GRID DATA project. 

Authors are with Department of Electrical and Computer Engineering, Texas A\&M University, College Station, TX, USA, 77843 (email: \{txu, hanyueli, abirchfield, overbye\}@tamu.edu). }

}

\IEEEoverridecommandlockouts
\IEEEpubid{\makebox[\columnwidth]{978-1-7281-0407-2/19/\$31.00~\copyright2019 IEEE \hfill} \hspace{\columnsep}\makebox[\columnwidth]{ }}
\maketitle
\IEEEpubidadjcol

\begin{abstract}
Big data analytic applications using phasor measurements help improve the situation awareness of grid operators to better operate and control the system. Phasor measurement unit (PMU) data from actual grids is viewed as highly confidential and is not publicly available to researchers and educators. This paper develops a methodology to synthesize PMU data that can be accessed and shared freely, with a focus on input data preparation.  Time series of demand- and generation-side input data are generated using public data from different utilities and statistical analysis methods. Detailed load dynamics modeling is also performed in this paper to extend the synthetic electric network models. 

\end{abstract}


\IEEEpeerreviewmaketitle

\section{Introduction}
\IEEEPARstart{P}{ower} system operation data can be obtained via multiple technologies and sources. Supervisory control and data acquisition (SCADA) is an industrial control system built to process, gather and monitor power system operation data \cite{SCADAandSmartGrids}. Implementations of phasor measurement units (PMUs) generate large amounts of time-synchronized data measured at a higher speed (e.g., thirty observations
per second), compared with traditional SCADA measurements (e.g.,one observation every several seconds) \cite{NASPI}. Big data analytics using those measurements have been applied to various diagnostics and decision support applications in power systems \cite{bigdatarole, FNET} to improve the security, reliability and efficiency of large-scale interconnected power systems. 

However, due to security concerns, the measurement data from actual grids is highly confidential, and so researchers have limited access to those data and even actual grid models, which limits big data researches and applications in the power community. Therefore, the research goal of this work is to create synthetic power system operation data using those public large-scale synthetic networks, which can be freely shared and used for various applications. 

Researchers in \cite{Synthetic1, Synthetic2, SyntheticCost, SyntheticDynConf, SyntheticDynTrans} used public data and statistics obtained from actual system models to build several large-scale synthetic electric networks that mimic the functional and structural features of actual systems without revealing any confidential information. This paper will address the applications of those large-scale network models for synthesizing time-synchronized phasor measurements primarily from two perspectives: time-varying input data and detailed load dynamic models. Time-dependent renewable energy outputs and electricity demands are the key driver of variations observed in phasor measurements. Since geographic information is also available in those synthetic networks, inclusion of those input data will consider both temporal and spatial correlations among different sites. The WECC composite load model - CMPLDW - with different parameter settings will be assigned to load buses, according to their residential/commercial/industrial (RCI) ratios.  This paper uses - ACTIVSg2000 - a 2000-bus case on the footprint of the Electric Reliability Council of Texas (ERCOT) as an example to illustrate the process \cite{SyntheticWebsite}, while the proposed method is general enough for other power system test cases and different time periods with various duration.

\section{Demand-Side Modeling and Input Data Creation Process}
This section first reviews the hourly load modeling procedure, through which each bus will obtain hourly load data for its residential, commercial and industrial classes, respectively. Statistical analysis performed on historical load data in 1-minute resolution will be applied to assign typical load variation patterns to each bus. According to the pre-defined RCI ratio, dynamic model components will be determined statistically for each bus.

\subsection{Overview of Hourly Load Modeling}
A bottom-up aggregation method is developed to create hourly load time series for the synthetic grid, where each load bus has a unique load profile \cite{SyntheticLoadModeling}. Among three levels of load hierarchy - building load, bus load, and area/system load, lower-level loads are aggregated to build the higher level load.

\begin{figure}[htb]
	\centering
	\includegraphics[width=3.3in]{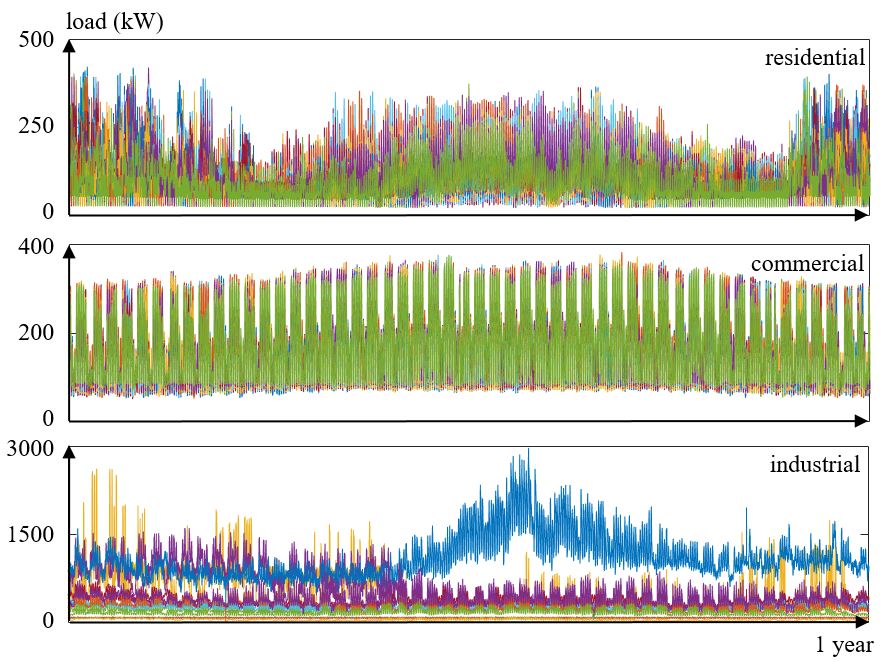}
	\caption{Building-level load time series (residential, commercial and industrial)}
	\label{figRCIPrototype}
\end{figure}

Building loads are generated under residential, commercial and industrial sectors, any of which have several prototypical load time series built from building-level energy consumption simulation data \cite{EIA861}.  Those building load time series represent load sector typical behaviors and potential variations. In the 2000-bus test case, residential, commercial and industrial sectors have 61, 60 and 12 prototypical building load time series, respectively. As shown in Fig.\ref{figRCIPrototype}, building load time series under the same load sector differ based on load size, weather region, and temporal behaviors. 

All prototype building-level loads are tagged with geographic coordinates. To aggregate from building loads to bus load, each bus is assigned with the geographically closest prototypical building load time series data from each of the residential, commercial and industrial sectors. The residential, commercial and industrial sales values \cite{ERCOTLoad} are used to determine the electric utility companies’ load class ratios. Given the geographic information of utility companies and synthetic grid buses, each bus is assigned to its closest electric utility company, inheriting its RCI ratio (the class ratio of electricity consumption from those sectors). 

The assigned prototypical building load time series are combined based on the optimal weights determined from a least square optimization problem in \cite{SyntheticLoadModeling}, where the peak of the weighted building load sum from three sectors should be close to the bus peak requirement from the base case, and the ratio of residential, commercial and industrial load satisfies the RCI ratio constraints. 

\begin{figure}[htb]
	\centering
	\includegraphics[width=3.3in]{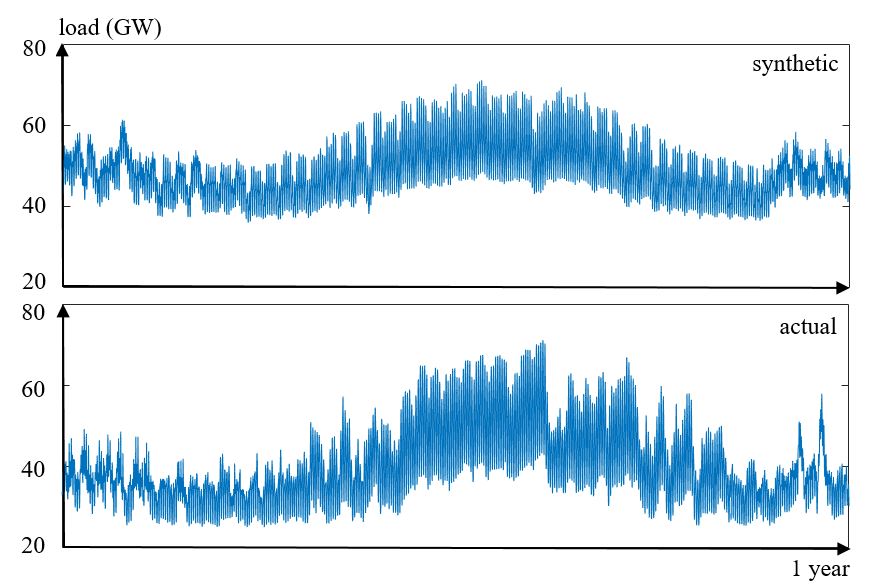}
	\caption{Comparisons of 2000-bus case and ERCOT system load time series}
	\label{fighourlyload}
\end{figure}

The summation of all the bus level load time series aggregate to the area/system level load. System level synthetic load time series exhibit similar characteristics compared to real electric power system of the same size. Fig.\ref{fighourlyload} shows 2000-bus test case system load, and ERCOT system load [5] in 2016.

\subsection{Load Time Series with Higher Time Resolution}

Running dynamic simulations requires time series input data with finer time resolution. Historical minutely load data \cite{CEPSMinuteLoad} is fully utilized in this paper to obtain characteristic patterns of time-dependent load variations.  Due to the time-dependent nature of human activities and weather conditions, electricity demand may establish distinct variation patterns at different periods of a day. Thus, for any simulation hour $h$, this paper collects load data of the specific hour from all days within the same season. Denote $[L_{d,1},L_{d,2},...,L_{d,m},...,L_{d,M}]^T$ as the $d^{th}$ load data in minutely time interval. For meaningful comparisons among load data obtained from different days, a scaling procedure is applied to process the collected data:
\begin{equation}
\hat{L}_{d,m}=\frac{L_{d,m}}{\frac{L_{d,M}-L_{d,1}}{M-1}\cdot (m-1)+L_{d,1}}, \text{for }\forall d,m
\end{equation}

\begin{figure}[htb]
	\centering
	\includegraphics[width=3.3 in]{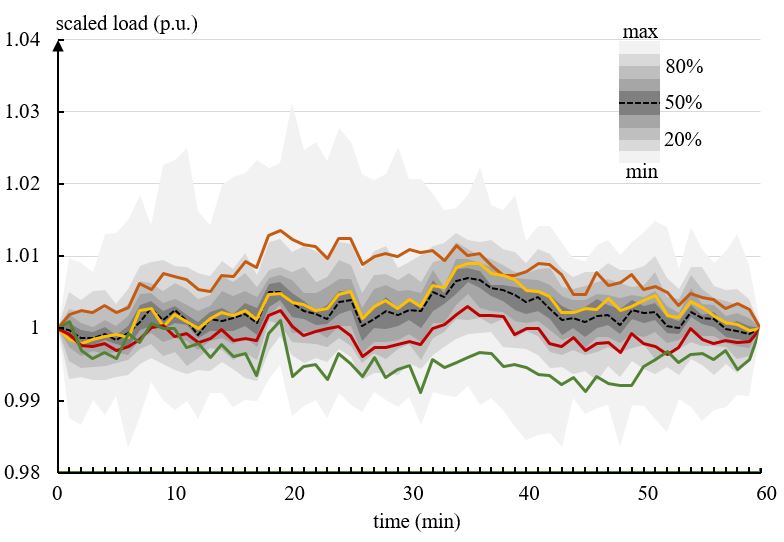}
	\caption{Percentile visualization and characteristic patterns of the scaled load data set}
	\label{figminuteloadpatterns}
\end{figure}

Since clustering techniques have been widely used for feature extraction and pattern analysis, this paper applies the \textit{K-means} clustering algorithm \cite{ClusteringBook} to obtain four characteristic patterns of the scaled load data set, as displayed in Fig.\ref{figminuteloadpatterns}. For references, Fig.\ref{figminuteloadpatterns} also visualizes the percentile plots of the scaled load data set.  Fig.\ref{figminuteloadpatterns} clearly shows that such a scaling procedure is able to capture the minute-by-minute load variations within each simulation hour, and the clustering results explicitly represent salient characteristics of different load variation patterns.

The key step for creation of load time series with finer time resolution is to assign the obtained minutely load patterns to a pre-defined hourly load curve  [$\tilde{L}_h,\tilde{L}_{h+1}$]. Consider a power grid with pre-defined $Z$ zones, each of which includes a group of buses with similar weather conditions and other demand-related factors. From a statistics point of view, the expected number of zones assigned with the $k^{th}$ minutely load pattern is $Z_k=\text{round}(Z\cdot p_k)$, where $p_k$ is the probability of the $k^{th}$ cluster. In addition, zones assigned to the same load pattern are more likely to be geographically close to each other, while patterns with more significantly different load variations are more possible to be assigned to zones with larger distances. To achieve so, we define a comparison metric $q_{z,z',k,k'}=\frac{D_{z,z'}}{dist_{k,k'}}$\footnote{$dist_{k,k'}$ is the Euclidean distance between two clusters $k,k'$ and we define $dist_{k,k}$ to be a sufficiently small positive number.}, and formulate the following problem to complete the assignment process: 

\begin{subequations} \label{eq_assignment_all}
\begin{align}
\label{eq_assignment_ob}
& \underset{c_{z,k}\forall z,k}{\text{max}}
& & -\sum_{z<z',k,k'}^{}\frac{\zeta_{z,z',k,k'}}{\zeta_{sum}}\lg(\frac{\zeta_{z,z',k,k'}}{\zeta_{sum}})\\ 
\label{eq_assignment_sumzZ}
& \text{\textit{s.t.}}
& & \sum_{z}^{}c_{z,k}=Z_k, \text{ }\forall k \\ 
\label{eq_assignment_sumz1}
& & & \sum_{k}^{}c_{z,k}=1, \text{ }\forall z \\ 
\label{eq_assignment_bi}
& & & c_{z,k} \in \{0,1\}, \text{ }\forall z,k \\ 
\label{eq_assignment_zeta}
& & & \zeta_{z,z',k,k'}=c_{z,k}c_{z',k'}q_{z,z',k,k'}, \text{ }\forall z<z',k,k' \\ 
\label{eq_assignment_zetasum}
& & & \zeta_{sum}=\sum_{z<z',k,k'}^{}\zeta_{z,z',k,k'} 
\end{align}
\end{subequations}

In constraint (\ref{eq_assignment_bi}), $c_{z,k}=1$ indicates that the $k^{th}$ load pattern is assigned to zone $z$. Constraint (\ref{eq_assignment_sumzZ}) defines that each cluster $k$ will be assigned to exactly $Z_k$ zones. Each zone can be assigned with one and only one load pattern as defined in constraint (\ref{eq_assignment_sumz1}). In (\ref{eq_assignment_zeta}), $\zeta_{z,z',k,k'}$ can be either a positive value or zero. In other words, $\zeta_{z,z',k,k'}$ is non zero only when zone $z$ is assigned with the $k^{th}$ load pattern ($c_{z,k}=1$) and zone $z'$ is assigned with the $k'^{th}$ load pattern ($c_{z',k'}=1$). Equivalently, $\zeta_{sum}$ is the sum of all non-zero $\zeta_{z,z',k,k'}$. Overall, the proposed formulation (\ref{eq_assignment_all}) aims to find an assignment such that ratios of the zone-pair distance to their corresponding cluster-pair distance are as close as possible for all zone pairs $z,z'$. Solutions to problem (\ref{eq_assignment_all}) can be obtained using any proper optimization techniques, which is outside the scope of this paper and thus won't be addressed here.

\begin{figure}[htb]
	\centering
	\includegraphics[width=3.3in]{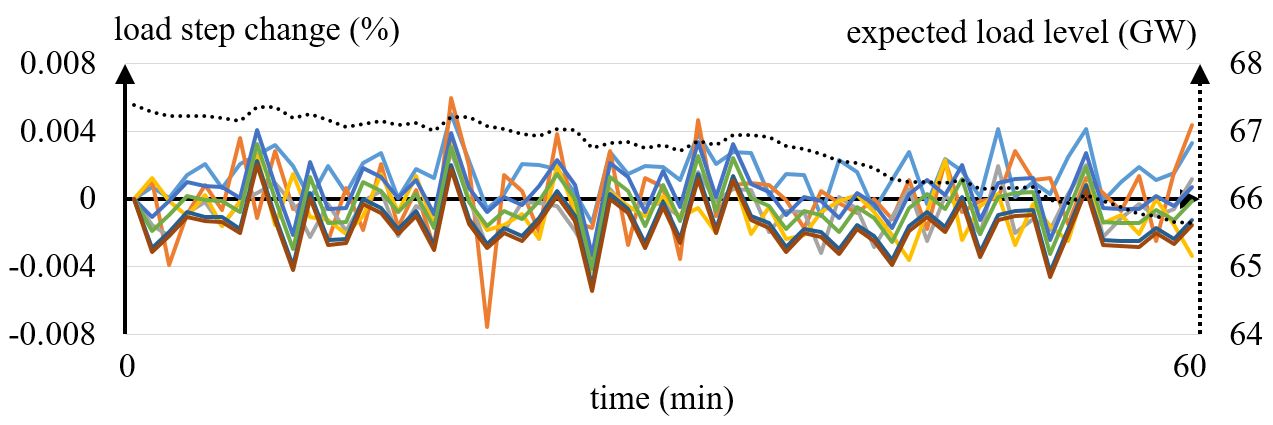}
	\caption{Load step change in percentage for eight weather zones and the expected system load (dotted line)}
	\label{figloadstepchange}
\end{figure}

Consider a zone $z$ assigned with $k^{th}$ load variation pattern $[\hat{l}_{k,1},...,\hat{l}_{k,m},...,\hat{l}_{k,M}]^T$. A re-scaling process is proposed to combine the obtained minutely load pattern $[\hat{l}_{k,1},...,\hat{l}_{k,m},...,\hat{l}_{k,M}]^T$ and the hourly load curve [$\tilde{L}_h,\tilde{L}_{h+1}$]:
\begin{equation}
\tilde{L}_{h,m}=(\frac{\tilde{L}_{h+1}-\tilde{L}_h}{M-1}\cdot (m-1)+\tilde{L}_h)\cdot \hat{l}_{k,m}, \text{for }\forall h,m
\end{equation}
By applying the re-scaling procedure to all loads, we obtain geographically and temporally correlated load curves in minutely time interval for all buses. Fig.\ref{figloadstepchange} shows zonal load step changes in percentage and the expected system load. For additional modeling purposes, such as forecast errors,  statistically random terms can be added to make the load profile more realistic. We note that, although this section focuses on the simulation hour $h$, the proposed method is also applicable to other hours of different days.

\subsection{Modeling Dynamics for Load with Different RCI Ratios}
Dynamic load modeling is essential for system transient stability studies \cite{IEEELoad1,IEEELoad2}. The load time series developed in previous steps can be viewed as the ideal demand trend over time, but a realistic simulation requires a proper representation of load dynamics. Historically the commonly used load dynamic models were constant impedance. However, the report by \cite{NERCLoadComposition1} pointed out that dynamic responses should reflect the actual behavior of aggregate loads. An up-to-date load dynamic model is an important factor to improve the accuracy of grid simulations \cite{LBNLLoadComposition}.

The Western Electricity Coordinating Council (WECC) load modeling task force has devoted the past years on development of a
composite load model, named as WECC Dynamic Composite Load Model (CMPLDW in PLSF; CMLDBLU1 in PSS/E) \cite{PNNLCMPLDW}. This model is designed to represent behaviors of different end-user components \cite{MotorGroups, NERCLoadComposition2,CMPLDWKQ}:
\begin{itemize}
\item Motor A: three-phase induction motors operating under
constant torque (\textit{i.e.} compressor motors used in large air-conditioners and refrigerators);
\item Motor B: three-phase induction motors with high inertia
operating under speed-dependent torque (\textit{i.e.} fan motor);
\item Motor C: three-phase induction motors with low inertia
operating under speed-dependent torque (\textit{i.e.} pump motor);
\item Motor D: single-phase induction motors (\textit{i.e.} heat pumps);
\item Electronic Load: electronics-driven electricity consumer (\textit{i.e.} personal computers, LED televisions);
\item Static Load: a frequency-dependent ZIP model (\textit{i.e.} resistive heating, cocking).
\end{itemize}

This work adopts the CMPLDW for modeling load dynamics. Table \ref{table_SampleLoadComposition} summarizes the adopted dynamic load composition \cite{MISORCItoLoad} for different periods of a day. The assignment process refers to Table \ref{table_SampleLoadComposition} and computes appropriate dynamic load component percentages for each load using its RCI ratio. Default parameters published by the North American Electric Reliability Corporation(NERC) Load Modeling Task Force are used for all CMPLDW models.

\def\arraystretch{1.3}%
\begin{table}[htb]
\caption{Sample Dynamic Load Compositions during Different Periods}
\begin{center}
\begin{tabular}{|c|c||c|c|c|c|c|c|}
\hline
\multirow{3}{*}{Period} & \multirow{3}{*}{Class} &
 \multicolumn{6}{c|}{Dynamic Load Component Percentage (\%)}  \\
\cline{3-8}
  &   &
 \multicolumn{4}{c|}{Motor}  &  \multicolumn{2}{c|}{Load} \\
\cline{3-8}
 & &  A & B & C & D & Electronic & Static \\
\hline
\multirow{3}{*}{Peak} & RES & 8 & 7 & 2 & 34 & 15 & 34\\
& COM & 12 & 10 & 4 & 25 & 18 & 31\\
& IND & 13 & 22 & 16 & 0 & 27 & 22\\
\hline
\multirow{3}{*}{Shoulder} & RES & 8 & 7 & 2 & 25 & 19 & 39\\
& COM & 12 & 10 & 4 & 20 & 23 & 31\\
& IND & 13 & 22 & 16 & 0 & 27 & 22\\
\hline
\multirow{3}{*}{Light} & RES & 10 & 8 & 2 & 0 & 40 & 40\\
& COM & 12 & 10 & 4 & 5 & 38 & 31\\
& IND & 13 & 22 & 16 & 0 & 27 & 22\\
\hline
\end{tabular}
\label{table_SampleLoadComposition}
\end{center}
\end{table} 

Through implementing the three steps proposed in this section, a non-conforming load profile with non-uniform dynamic load compositions is obtained. Such a profile captures the geographic and temporal correlations among different sites, and is capable to be used for transient stability studies using large-scale network models.

\section{Supply-Side Modeling and Input Data Creation Process}
Time-dependent variations and uncertainty exist in both demand and supply sides. This section aims to build time-varying wind power profiles, by starting with wind speed time series in 5-minute time resolution. Then,  analysis on historical secondly wind speed data is performed to obtain statistic characteristics of wind speed variations, which is used to generate synthetic wind speed variation terms in finer time resolution. A procedure is proposed to combine the wind speed measurements in 5-minute time interval and wind speed variation terms in finer time resolution. At last, a common wind power production model is adopted for converting wind speeds to wind farm outputs.

\subsection{5-min Wind Speed Time Series}
\begin{figure}[htb]
	\centering
	\includegraphics[width=3.3in]{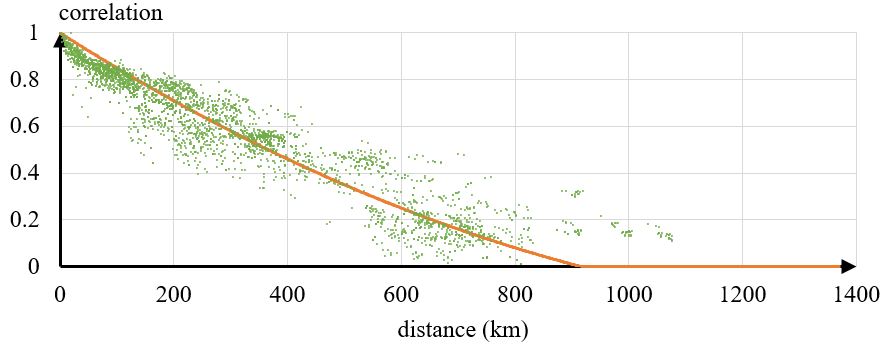}
	\caption{Correlations versus distances among wind farms}
	\label{fig5minwindcorrelation}
\end{figure}

Wind Integration National Dataset (WIND) Toolkit is a wind data set developed and maintained by the National Renewable Energy Laboratory (NREL) \cite{NRELWindPaper,NRELWindSite}. In this paper, simulated 5-min wind speed profiles for all synthetic wind farms in the test case are obtained from this toolkit, according to their geographic information - latitudes and longitudes. For further data analysis, we compute correlations among all synthetic wind farms and Fig.\ref{fig5minwindcorrelation} displays the relationship between correlations and site distances. A two-segment polynomial curve $\rho_{i,j}=\rho(D_{i,j})$ is also obtained to approximate such relationship. The collected 5-min wind speed time series form the basis for inclusion with time-dependent variation terms in finer time resolution, which will be discussed in the next step.

\subsection{Wind Speed Time Series with Finer Time Resolution}

Creation of wind speed time series in finer time resolution faces a key challenge - the inherent uncertainty, variability and intermittency features of wind energy. This subsection applies a set of secondly wind speed measurements from \cite{1HzWindSpeedSite} for statistical analysis. Equations (\ref{eq_windspeed_conversion}) are formulated to process one secondly data series $[W_{1},...,W_{s},...,W_{S}]^T$ over a period of a predefined duration. 

\begin{subequations} \label{eq_windspeed_conversion}
\begin{align}
\hat{W}_{s} & =W_{s}-(\frac{W_{S}-W_{1}}{S-1}\cdot (s-1)+W_{1}), \text{ }\forall s \\
\label{eq_windspeed_conversion_step2}
\hat{\hat{W}}_{s} & =\hat{W}_{s}-\hat{W}_{s-1}, \text{ }\forall s \geq 2
\end{align}
\end{subequations}

Here $\hat{\hat{W}}_{1}$ is defined to be zero. Clearly, the set $\{\hat{\hat{W}}_{s},\forall s\}$ over one period has a zero mean and a non-zero standard deviation. This paper uses normal distribution $\mathcal{N}(0,\sigma)$ to approximate the distribution of the set $\{\hat{\hat{W}}_{s},\forall s\}$. Thus, we repeat this process for secondly wind speed data set from each of all same-duration periods on a rolling basis. 
\begin{figure}[htb]
	\centering
	\includegraphics[width=3.3in]{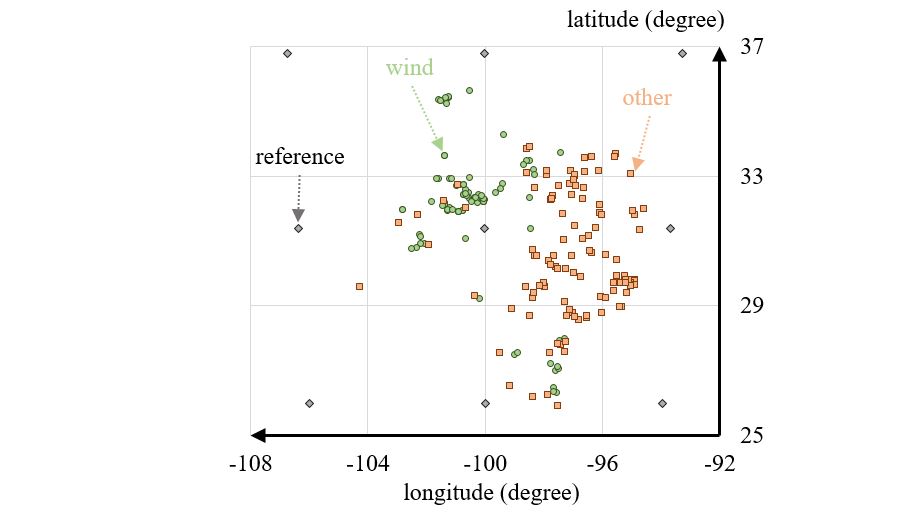}
	\caption{Geographic representation of generators and selected reference points}
	\label{figwindreference}
\end{figure}

The locations of wind and non-wind generators are shown in Fig.\ref{figwindreference}. In this paper, we consider a reference network with $N$ points uniformly distributed over the footprint. According to Fig.\ref{fig5minwindcorrelation}, two wind farm sites with a distance at least 600 km is viewed as trivially correlated. Thus, as shown in Fig.\ref{figwindreference}, those reference points are selected such that two neighboring sites in each East-West row and each North-South column have a distance of 600 km. For each reference point $n$, we statistically select a $\sigma \sim \psi(\cdot)$ and generate a variation set $[\hat{\hat{w}}_{n,1},...,\hat{\hat{w}}_{n,s},...,\hat{\hat{w}}_{n,S'}]^T$ over a 5-minute period from $\mathcal{N}(0,\sigma)$, where $\hat{\hat{w}}_{n,1}$ is defined to be zero.  Correspondingly, we can compute the wind speed variation terms $[\hat{w}_{n,1},...,\hat{w}_{n,s},...,\hat{w}_{n,S'}]^T$ by taking $\hat{w}_{n,s}=\hat{w}_{n,s-1}+\hat{\hat{w}}_{n,s}$ and $\hat{w}_{n,0}=0$, which can be viewed as an inversion of (\ref{eq_windspeed_conversion_step2}). As such, we can further build the wind speed variation terms  $[\hat{v}_{e,1},...,\hat{v}_{e,s},...,\hat{v}_{e,S'}]^T$ for each synthetic wind farm $e$ over a 5-minute period using those reference points:
\begin{equation}
\hat{v}_{e,s} =\sum_{n}\frac{\pi_{e,n}\hat{w}_{n,s}}{ \Omega_{e}}, \text{where } \Omega_{e}=\sum_{n}\pi_{e,n}, \forall e 
\end{equation}
$\pi_{e,n}$ is the weight factor for wind farm $e$ and reference point $n$. In this paper, the correlation value computed based on Fig.\ref{fig5minwindcorrelation} is used as the weight factor: $\pi_{e,n}=\rho_{e,n}=\rho(D_{e,n})$. 
 
Suppose that $V_{e,\tau}$ and $V_{e,\tau+1}$ are wind speeds from Section III.A - two end points of a 5-minute time period. The wind speed vector $[v_{e,1},...,v_{e,s},...,v_{e,S'}]^T$ for each synthetic wind farm $e$ over a 5-minute period can be computed by:
\begin{equation}
	\label{eq_fromreference_tosite_step2}
	v_{e,s}  =(\frac{V_{e,\tau+1}-V_{e,\tau}}{S'-1}\cdot (s-1)+V_{e,\tau})+\hat{v}_{e,s}, \forall s\in (1,S')
\end{equation}
where $v_{e,1}$ and $v_{e,S'}$ are set to be equal to  $V_{e,\tau}$ and $V_{e,\tau+1}$, respectively. Overall, equation (\ref{eq_fromreference_tosite_step2}) combines the wind speed data $[V_{e,\tau},V_{e,\tau+1}]^T$ in the 5-minute time interval (from Section III.A) with the 5-minute-duration wind speed variation vectors $[\hat{v}_{e,1},...,\hat{v}_{e,s},...,\hat{v}_{e,S'}]^T$ in a finer time interval (from Section III.B), to formulate the wind speed data $[v_{e,1},...,v_{e,s},...,v_{e,S'}]^T$ in a finer time interval for a 5-minute period. Such a procedure is then repeated to generate multi-site wind speed profile over a longer period (e.g., one hour or a day).

\subsection{Wind Power Production Model}
This paper adopts standard terminology \cite{WindModelBook} for wind power production model: cut-in ($v_{cutin}$), rated ($v_{rated}$), and furling wind speed ($v_{furl}$) of a wind turbine. The following continuous function \cite{WindModelGarcia} is used to model the power output $p(\cdot)$ (in p.u.) of the wind turbine, as a function of wind speed $v$:
\begin{subequations}
	\begin{align}
	p(v) & = \left\{
	\begin{array}{ll}
	1-(1+\exp(\frac{v-v_{mid}}{\alpha_1}))^{-1}, \text{ } 0\leq v \leq v_{lim}\\
	1-\alpha_2 (v-v_{lim})^{\alpha_3}, \text{ } v_{lim}< v \leq v_{furl} \\
	0, \text{ otherwise} 
	\end{array}
	\right.\\
	\alpha_1 & = (v_{rated}-v_{mid})(\ln(\frac{\beta}{1-\beta}))^{-1} \\
	\alpha_2 & = (v_{furl}-v_{lim})^{-\alpha_3}
	\end{align}
\end{subequations}
$v_{mid}$ is the wind speed where the wind power output is expected at half the rated value. Shape parameter $\alpha_1$ is tuned by changing $\beta$ (typically close to 1) such that the power output at the rated speed can almost achieve the rated power. Shape parameters $\alpha_2$ and $\alpha_3$ are selected to model a drop-off power output function after the wind speed exceeds the limiting value $v_{lim}$ but still within the $v_{furl}$. Parameter values from \cite{WindModelGarcia} are modified for uses in this paper with $\beta=0.95$.

\begin{figure}[htb]
	\centering
	\includegraphics[width=3.3in]{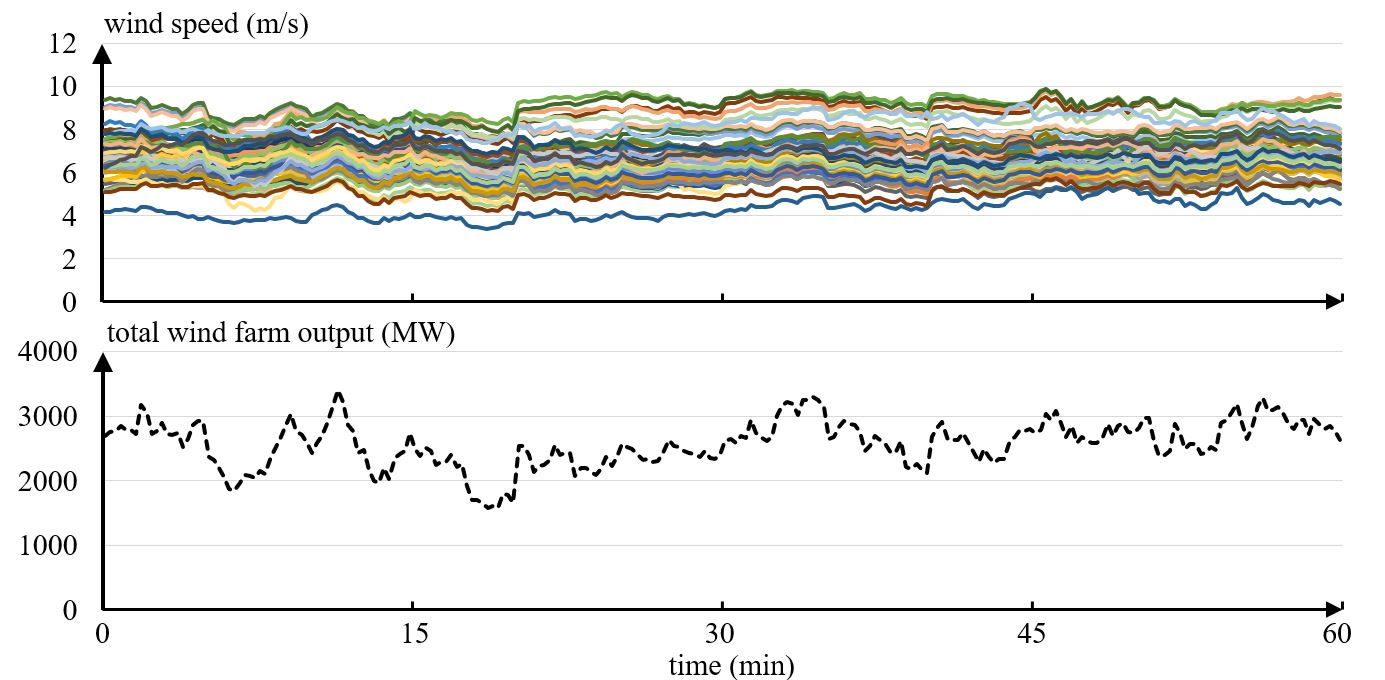}
	\caption{Multi-site wind speeds and system total wind power}
	\label{figwindspeedandtotaloutput}
\end{figure}

As a summary, Fig.\ref{figwindspeedandtotaloutput} summarizes the simulated wind speeds at multiple wind farms in 15-s time resolution for one hour and the corresponding total wind power output. Those curves show that the proposed method is capable to generate correlated wind speed and output profiles that capture the uncertainty and variability nature of wind resources.

\section{Synthetic PMU Data Creation Example}
A simulation of 10-minute with half cycle time step is performed on the ACTIVSg2000 system to create the example synthetic PMU data. PowerWorld Dynamic Studio, an interactive environment for transient stability level simulation, is used \cite{PWDS}. 

For consistent input data granularity, the 1-min interval load data is superimposed with Gaussian noise of 1\% standard deviation every 15 seconds.  The load and wind variation data is then formatted to be a time series binary (tsb) file as the simulation input. System control actions, such as automatic generation control, and economic dispatch, are also taken into consideration for the realistic creation of synthetic PMU data. Generators in the synthetic system are re-dispatched every 15 minutes to mimic real time operations from the control rooms.  

\begin{figure}[h]
	\centering
	\includegraphics[width= 3.3in]{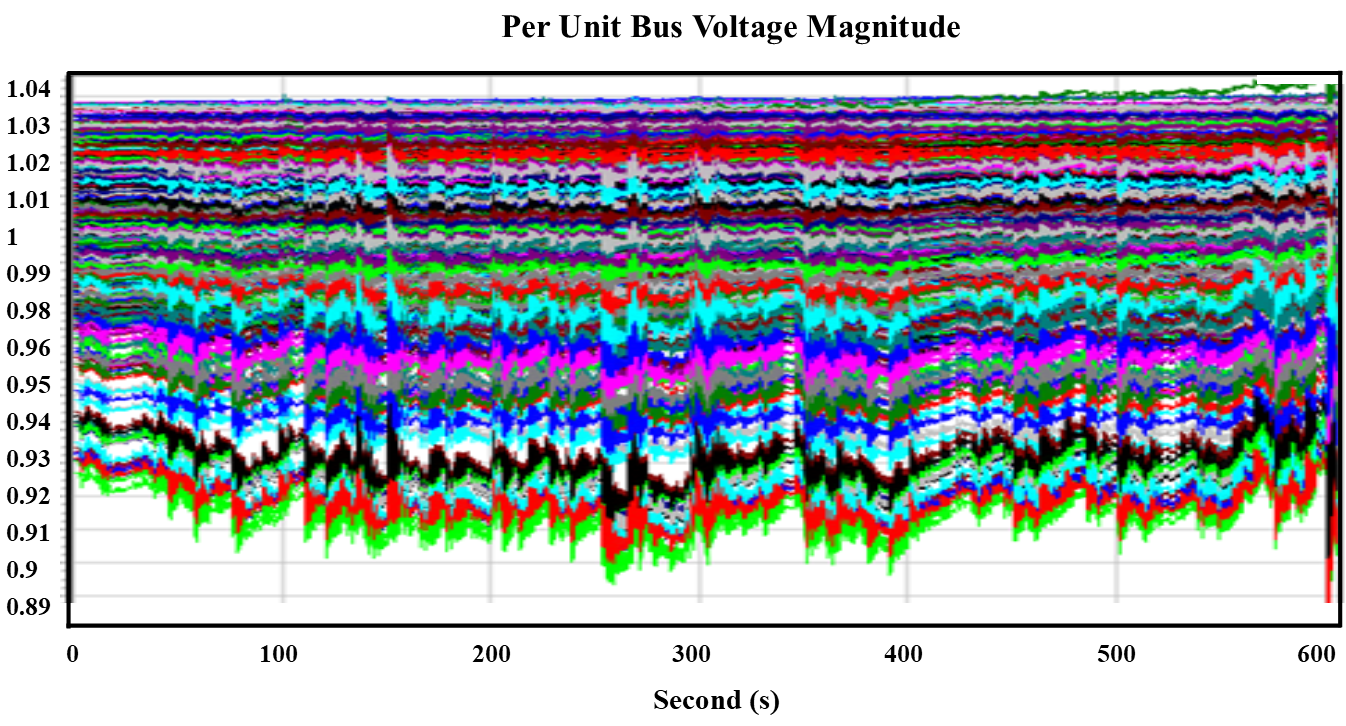}
	\caption{Per unit bus voltage magnitude}
	\label{fig:result}
\end{figure}

Figure \ref{fig:result} shows the created PMU data of voltage magnitude for all the buses in the ACTIVSg2000 synthetic system. It is observed that the synthetic PMU voltage data reflects the variations in load, renewable generation and generator controls. 

\section{Conclusion and Future Works}
This paper proposed a methodology of creating PMU data from large-scale synthetic electric systems, with a focus on the input data preparation. The creation of time-varying input data of both demand and generation, as well as the detailed dynamic load models are discussed. An example of generating synthetic PMU data using created time-varying inputs is also provided. The produced synthetic PMU data can be accessed and distributed freely without confidentiality concerns. 

For future work, the post-modification and adaptation of PMU signals, as well as statistical validation of the created data are critical to improve the realism of the synthetic PMU data.

\section*{Nomenclature}
\addcontentsline{toc}{section}{Nomenclature}
\begin{IEEEdescription}[\IEEEusemathlabelsep\IEEEsetlabelwidth{$V_1,V_2,V_3$}]
	\item[ ] \textbf{Section II}
	\item[$d$] index of day from 1 to $D$
	\item[$m$] index of minute from 1 to $M$
	\item[$L_{d,m}$] the measured load level at minute $m$ of the specific hour for day $d$ (MW) 
	\item[$\hat{L}_{d,m}$] the scaled load level at minute $m$ of the specific hour for day $d$ (p.u.) 
	\item[$\hat{l}_{k,m}$] the scaled load level at minute $m$ in the $k^{th}$ load variation pattern (p.u.) 
	\item[$k$] index of cluster from 1 to $K$
	\item[$p_k$] probability of the $k^{th}$ cluster
	\item[$z$] index of zone from 1 to $Z$

	\item[$q_{z,z',k,k'}$] comparison metric defined as $\frac{D_{z,z'}}{dist_{k,k'}}$
	\item[$c_{z,k}$] status indicator equal to 1 if and only if the $k^{th}$ load pattern is assigned to zone $z$
	
	\item[$\zeta_{z,z',k,k'}$] defined as $c_{z,k}c_{z',k'}q_{z,z',k,k'}$
	\item[$\zeta_{sum}$] defined as $\sum_{z<z',k,k'}^{}\zeta_{z,z',k,k'} $

	\item[$\tilde{L}_h,\tilde{L}_{h+1}$] two end points of the hourly load curve (MW) 
	
	\item[$\tilde{L}_{h,m}$] the re-scaled load level at minute $m$ of the specific hour $h$ (MW)

	\item[ ] \textbf{Section III}
	\item[$\rho_{i,j}$] correlation between two sites $i,j$
	\item[$s$] index of second from 1 to $S$
	\item[$W_{s}$] the measured wind speed at second $s$ (m/s) 
	\item[$\hat{W}_{s}$,$\hat{\hat{W}}_{s}$] variation terms defined in equation (\ref{eq_windspeed_conversion})
	
	\item[$\sigma$] the standard deviation of $\hat{\hat{W}}_{s}$
	\item[$\psi(\cdot)$] approximated probability distribution function for $\sigma$
	
    \item[$n$] index of reference point from 1 to $N$
	\item[$\hat{\hat{w}}_{n,s}$] sampled variation term
	\item[$\hat{w}_{n,s}$] defined as $\hat{w}_{n,s}=\hat{w}_{n,s-1}+\hat{\hat{w}}_{n,s}$ and $\hat{w}_{n,0}=0$

	\item[$e$] index of wind farm from 1 to $E$
	\item[$\hat{v}_{e,s}$] computed wind speed variation term at farm $e$
	\item[$v_{e,s}$]  wind speed at farm $e$ (m/s) 
	
	\item[$\pi_{e,n}$]  weight factor for wind farm $e$ and reference point $n$ with $\Omega_{e}=\sum_{n}\pi_{e,n}$
	\item[$V_{e,\tau}, V_{e,\tau+1}$] wind speeds - two end points of a 5-minute time period.
	
	\item[$v_{cutin}$]  cut-in wind speed (m/s) 
	\item[$v_{mid}$]   wind speed where the wind power output is expected at half the rated value (m/s) 
	\item[$v_{rated}$]  wind speed where the wind power output is expected at the rated value (m/s) 
	\item[$v_{lim}$]  limiting wind speed (m/s) 
	\item[$v_{furl}$]  furling wind speed (m/s) 
	\item[$\alpha_1,\beta$] shape parameters for wind power production model when $0\leq v \leq v_{lim}$
	\item[$\alpha_2,\alpha_3$] shape parameters for drop-off power output function when $v_{lim}< v \leq v_{furl} $

\end{IEEEdescription}

\bibliographystyle{IEEEtran}
\bibliography{bibi}

\end{document}